# A unified molecular level mechanism for the universal α- and Johari-Goldstein β-relaxations in glassformers

Y.N Huang [a,b*], J.L. Zhang [a,b], L.L. Zhang [a] and L.N. Wang [a]

*(a) College of Physics and Electronic Information and Institute of Condensed Matter Physics and Design, Ili Normal University, Yining, 835000*

*(b) Department of Physics and National Lab of Solid State Microstructures, Nanjing University, Nanjing 210093*

Abstract

We presented that the relaxation of *n* coupling molecules in a molecular string exhibits *n* individual relaxation modes (RMs), each mode being characterized by a definite relaxation time and amplitude according to the string model. The *n* RMs behaving a single relaxation at high temperature, evolves to two relaxation species, at low temperature, with different temperature dependences for the respective relaxation times and amplitudes. Since the characteristics of the two relaxation species are in agreement with those exhibited by the universal α- and Johari-Goldstein (JG) β-relaxations in glass dynamics, we provided a unified molecular level mechanism for these two processes.

1. Introduction

Glass transition is a fundamental property of the condensed matter [1-4]. Though some quite successful phenomenological [5-8] and coarse-grained theories [9-11] have been formulated to describe the glass transition, none of the microscopic mechanisms proposed to explain at the molecular level this phenomenon has received widely acceptance, thus remaining a central unresolved issue in the physics of condensed matter [12]. However, it cannot be denied that some of the proposed microscopic models, potentially or in part, describe the key characteristics of the transition, such as the glass structure and the glass dynamics including the universal α relaxation, the Johari-Goldstein (JG) β-process, the Boson peak etc. [13-15]. Among these models, the string model [16] has the advantage of providing a unified picture of the glass structure for both monomeric and macromolecular glasses, i.e. the random packing of strings or macromolecules in space. In this paper we use the string model to investigate, at the molecular level, the microscopic

---

mechanism of the universal α- and JG β-relaxations in glass dynamics.

## 2. Molecular string model

Besides describing the single molecular motions of Debye's mean-field theory approximation [17], the string model represents the collective units of the structure and dynamics in glassformers by coupled molecular strings randomly distributed in space, characteristics which are widely observed in well-designed experiments [18-19], analog simulations [20] and molecular dynamics simulations [21]. Furthermore, the model assumes relative weak interactions between the strings, compared with the intra-string ones. The Hamiltonian describing orientational motions in Debye's theory [17], is the well-known sin-Gordon potential, $H_0 = \frac{V_0}{2}\sum_i \{1 - \cos[2(\varphi_i - \varphi_i^0)]\}$, where $V_0$ is the barrier height between the double-wells, $\phi_i$ ($0 \leq \phi_i < \pi$) is the orientational angle of the i$^{th}$ molecule in the system, and $\phi_i^0$ is the referenced angle in the range $[0,\pi]$. Translational movements are described by the $\phi^{(4)}$ potential, $H_0 = V_0 \sum_i \left[ -2\left(\frac{x_i}{d}\right)^2 + \left(\frac{x_i}{d}\right)^4 \right]$, where $V_0$, $x_i$ and $d$ are, respectively, the barrier height between the double-wells, the displacement of i$^{th}$ molecule in the system and the well width. Furthermore, molecular jumping processes between the double-wells ($e^{-V_0/k_B T} \ll 1$, where $k_B$ is the Boltzmann's constant, and $T$ the absolute temperature) are equivalent to those between two states, $\sigma_i = \pm 1$, with jumping rate $v_0 e^{-V_0/k_B T}$ ($v_0$ is the vibration frequency of molecules in the wells). Taking into account these principles, the string model represents intra-string interactions by a finite one dimensional Ising model $H_1 = -V \sum_m \sum_{i=1}^{n-1} \sigma_i^{mn} \sigma_{i+1}^{mn}$ [17], whereas the inter-string interactions are represented by a random Ising interacting model $H_2 = \frac{V'}{2} \sum_m \sum_{i=1}^{n} \sum_{m' \neq m, i'}^{nn(i)} \sigma_i^{mn} \sigma_{i'}^{m'n'} \Theta_{ii'}^{mm'}$, where $V$ and $V'$ are the corresponding interaction constants, independent of temperature. The symbol $\sigma_i^{mn} = \pm 1$ denotes the two states of the i$^{th}$ molecule of a string (numbered $m$ in the system) containing $n$ molecules (called $n$-string hereafter), $nn(k)$ represents the nearest number of molecules surrounding the i$^{th}$ molecule, and $\Theta_{ii'}^{mm'}$ is a random number in the range $[-1, 1]$. In the case of orientational motions, $\Theta_{ii'}^{mm'}$ is the direction cosine of the molecule $i$, in the $n$-string, with the molecule $i'$ in the $n'$-string. For the Debye's theory only provides a jumping time for the molecules between the double-wells $\tau_0 = v_0^{-1} e^{V_0/k_B T}$, the effective Hamiltonian of the string model describing the coupled orientational and translational jumping motions is given by,

$$H_e = -V \sum_{m}\sum_{i=1}^{n-1} \sigma_i^{mn} \sigma_{i+1}^{mn} + \frac{V'}{2} \sum_{m}\sum_{i=1}^{n} \sum_{m'\neq m,i'}^{nn(i)} \sigma_i^{mn} \sigma_{i'}^{m'n'} \Theta_{ii'}^{mm'} \qquad (1)$$

This equation is the Hamiltonian of a partly random Ising model [17].

## 3. Model results and discussion

Molecular strings comprise the essential elements of the string model for both glass structure and dynamics. According to the intra-string interactions in Eq.(1), the relaxation equation for an individual $n$-string is [16],

$$\frac{d}{dt}\begin{bmatrix} \delta_1 \\ \delta_2 \\ \vdots \\ \vdots \\ \vdots \\ \delta_{n-1} \\ \delta_n \end{bmatrix} = -\frac{1}{\tau_0} \begin{bmatrix} 1 & 2u-1 & 0 & \cdots & 0 & 0 & 0 \\ w-1/2 & 1 & w-1/2 & \cdots & \vdots & 0 & 0 \\ \vdots & w-1/2 & 1 & \cdots & \vdots & \vdots & 0 \\ \vdots & \vdots & w-1/2 & \cdots & w-1/2 & \vdots & \vdots \\ 0 & \vdots & \vdots & \cdots & 1 & w-1/2 & \vdots \\ 0 & 0 & \vdots & \cdots & w-1/2 & 1 & w-1/2 \\ 0 & 0 & 0 & \cdots & 0 & 2u-1 & 1 \end{bmatrix} \begin{bmatrix} \delta_1 \\ \delta_2 \\ \vdots \\ \vdots \\ \vdots \\ \delta_{n-1} \\ \delta_n \end{bmatrix} \qquad (2)$$

where $\delta_i^{(n)}(t) \equiv p_i^{(n)}(t) - p_i^{(n)}(\infty), i = 1,\cdots,n$, $p_i^{(n)}(t)$ is the probability that the i[th] molecule of the $n$-string is in the state $\sigma_i^{mn} = 1$, and $p_i^{(n)}(\infty)$ is the relevant probability at equilibrium. Therefore, $\delta_i^{(n)}(t)$ is the departure of the property from equilibrium. $u \equiv (1 + e^{2V/k_B T})^{-1}$, and $w \equiv (1 + e^{4V/k_B T})^{-1}$.

For the sake of convenience, Eq.(2) will be called the string relaxation equation (SRE) whereas the square matrix on the right side of Eq.(2) will be named $M_{SR}$. At temperatures high enough, i.e. $V/k_B T \to 0$, $M_{SR}$ becomes a unit matrix and the SRE simplifies to the Debye relaxation equation [17]. On the other hand, at temperatures low enough, i.e. $V/k_B T \to \infty$, $M_{SR}$ is the Rouse-Zimm matrix, the SRE becoming the Rouse-Zimm relaxation equation [22]. Since the Debye and Rouse models are, respectively, the two most successful relaxation theories for monomeric and macromolecular glassformers, the string model can be considered, at least mathematically, a universal model that describes the relaxation dynamics of the amorphous condensed matter.

To solve the SRE let us assume a square matrix $[X]$ such that $[M_{SR}][X] = [X][D]$ where $[D]$ is a diagonal matrix. By defining a vector $[R] \equiv [X]^{-1}[\delta]$, then the SRE can be transformed to a diagonal form,

$$\frac{d}{dt}[R] = -\frac{1}{\tau_0}[D][R] \tag{3}$$

The solution of this equation is,

$$\begin{cases} R_j^{(n)}(t) = R_j^{(n)}(0) e^{-t/\tau_j^{(n)}} \\ \tau_j^{(n)} \equiv \tau_0 / D_{jj}^{(n)} \\ R_j^{(n)}(0) = \sum_{i=1}^{n} r_{ji} \\ r_{ji} \equiv [X^{-1}]_{ji}^{(n)} \delta_i^{(n)}(0) \end{cases}, \; i, j = 1, \cdots, n \tag{4}$$

Physically, the mathematical process translates the relaxation of the $n$ coupled molecules in an $n$-string to $n$ relaxation modes (RMs) $R_j^{(n)}(t)$, $j = 1, \cdots, n$, with definite relaxation time $\tau_j^{(n)}$ and amplitude $R_j^{(n)}(0)$. The spatial RMs distribution is expressed by $r_{ji}^{(n)}$ whereas $\delta_i^{(n)}(0), i = 1, \cdots, n$ is the initial condition of the SRE, which depends on the experimental method.

Broadband dielectric spectroscopy allows to measure the dynamics of molecular systems over more than ten decades, thus providing a useful experimental tool to obtain the α- and JG β-relaxations of supercoolod liquids in a wide frequency range [13-15]. In a step electric field, $E(t) = \begin{cases} E_0, t < 0 \\ 0, t \geq 0 \end{cases}$, and without losing generality for the results as well as in the linear response regime, i.e. $\mu_0 E_0 / k_B T \ll 1$ [17], the value of $\delta_i^{(n)}(0)$ is given by [23]

$$\delta_i^{(n)}(0) = \frac{\mu_0 E_0}{2 k_B T} \sum_{k=1}^{n} [\tanh(V/k_B T)]^{|k-i|} \cos \theta_k^{(n)}, i, k = 1, \cdots, n \tag{5}$$

where the permanent electric dipole moment of the molecule along the string contour line is $\mu_0$ and $\theta_k^{(n)}$ is the angle between $E_0$ and $\mu_0$ for the k$^{th}$ molecule in the string that depends on the spatial configurations of the string [16,23].

Figs.1 and 2 show Arrhenius plots for the relaxation time, $\tau_j^{(29)}$, and amplitude, $R_j^{(29)}(0)$, of the RMs of a linear 29-string ($\theta_k^{(29)} = 0, k = 1, \cdots, 29$). The calculations were carried out using Eqs.(2)-(5), and the numbering of the RMs from $j = 1$ to $n$ corresponding to $\tau_j^{(n)}$ appears from high to low values. Except for the first RM, $\tau_j^{(n)}$ and $R_j^{(n)}(0)$ for the other RMs have the same evolution with *T*.

Specifically, the relaxation time $\tau_j^{(n)}/\tau_0, j \geq 2$ firstly increases with decreasing temperature tending to a definite value, whereas $R_j^{(n)}(0), j \geq 2$ first increases reaching a maximum, and then almost exponentially decreases. In contrast, $\tau_1^{(n)}/\tau_0$ associated with the first RM always goes up, the low temperature asymptotic being $\tau_{1T}^{(n)} = \tau_0 e^{2V/k_BT}(n-1)/2$ [16], while $R_1^{(n)}(0)$ first increases and then tends to a definite value. The inset of Fig.2 presents the spatial distribution $r_{ji}^{(29)}$ for the first to the fifth RMs of the linear 29-string. The results indicate that except for the nearly uniform distribution of the first RM, the others modes oscillate in space, in such a way that the extreme points for each RM are equal to their numbering.

The results discussed above indicate that the $n$ RMs of an $n$-string can be specified as corresponding to two species, one called $R_a^{(n)}(t) = R_1^{(n)}(t)$, and the other $R_b^{(n)}(t) = \sum_{j=2}^{n} R_j^{(n)}(t)$. At high temperature, the two species have nearly the same $\tau_j^{(n)}$, thus comprising a single relaxation process more or less broader than the Debye relaxation, called $R_{ab}^{(n)}(t) = \sum_{j=1}^{n} R_j^{(n)}(t)$ here. With decreasing temperature, the single process splits into two individual relaxations. An obvious conclusion is that the loss peak of $R_b^{(n)}(t)$ in the frequency domain is wider than that of $R_a^{(n)}(t)$, but its activation energy is smaller.

In what follows, we analyze the influence of both the inter-string random Ising interaction and the string length distribution $g_n$ (which only depends on $V/k_BT$ and the effective coordination number, $z_e$, of a molecule) [16] on both the average relaxation times $\tau_a$ and $\tau_b$. The analysis will be extended to the study of the effect of both interaction and distribution on the relaxation functions $\varphi_a$ and $\varphi_b$ associated, respectively, with $R_a(t) \equiv \sum_n g_n R_a^{(n)}(t)$ and $R_b(t) \equiv \sum_n g_n R_b^{(n)}(t)$.

As shown elsewhere [16], molecular thermal fluctuations are strong at high temperature, and the net attraction energy between strings is zero. With decreasing temperature, the molecular occupation order of the first RM (inset of Fig.2) induces the ordering of surrounding strings, and consequently non-zero net attraction energy appears at a certain temperature By taking into account this precedent, the calculated temperature dependence of $\tau_a$ is $\exp\left[\dfrac{A + B\exp(C/T)}{T}\right]$, where $B$ is proportional to the net attraction energy. The value of $B$ changes from zero at high temperature to a value larger than zero at low temperature, whereas $A$ and $C$ are constants, independent of temperature. This result indicates the existence of a crossover from the high temperature Arrhenius relation to the low temperature super-Arrhenius one [16], which is nearly similar to the Vogel-Fulcher-Tammann (VFT) equation [24] that fits the temperature dependence of the relaxation times of α-absorptions. Moreover, the high temperature $\varphi_a$ is nearly an exponential function because of the string length distribution. However, at low temperature, the non-zero net interaction leads to an increasing difference of the relaxation times of the 1st RM corresponding to different values of $n$ and, consequently, $\varphi_a$ shows a strongly stretched exponential behavior. This is consistent with the Havriliak-Negami (HN) [25], Cole-Davidson (CD) [26] and Kohlrausch-Williams-Watts (KWW) [27] empirical laws, depending on the value of $z_e$ [16,23].

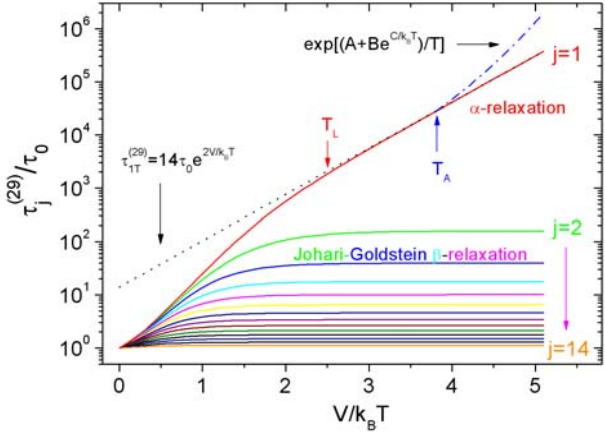

Fig.1 Relaxation times $\tau_j^{(29)}$ of 1st to 14th RMs of a 29-string versus reciprocal of temperature $1/T$. The black dot line is the exact solution of $\tau_1^{(29)}$ at low temperature limit and the blue dash dot line is the modified $\tau_1^{(29)}$ by the inter-string interaction according to the Ref.16.

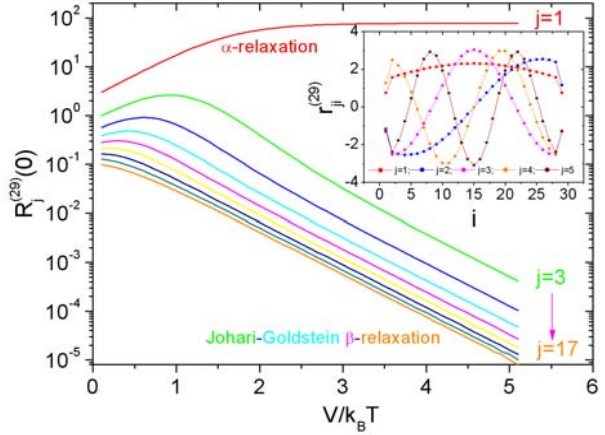

Fig.2 Relaxation amplitudes $R_j^{(n)}(0)$ of the 1st, 3rd, … and 17th relaxation modes (RMs) of the linear 29-strings versus the reciprocal of temperature $1/T$. Inset shows the spatial distribution $r_{ji}^{(29)}$ of 1st to 5th RMs.

For $R_b(t)$, the spatial amplitude oscillation of the RMs will greatly reduce the net attraction among surrounding strings, and it should be expected that $\tau_b$ will follow the Arrhenius relation. Meanwhile, the corresponding behavior of $\varphi_b$ in frequency domain is described by the Cole-Cole (CC) empirical law [28] because $R_b(t)$ contains many RMs even for a single string besides the string lengths distribution. The details will be published elsewhere [23].

The characteristics discussed above for $R_a(t)$ and $R_b(t)$ remind the universal α- and JG β-relaxations of glass dynamics. Experiments show that α- and JG β-relaxations are just different exhibitions of the responses of the same interacting molecules. At high temperature, they merge into a single relaxation, called αβ-relaxation, which splits again into the two processes if the temperature is lowered [13-15]. The string model predicts that the JG β-relaxation in the frequency domain is much wider than the α- relaxation, but its activation energy is smaller. The model also predicts that the average relaxation time $\tau_\alpha$ of the α-relaxation evolves from high temperature Arrhenius to low temperature VFT law [16]. The model also gives a good account of the crossover of the relaxation function from nearly exponential, at high temperature, to a low temperature stretched exponential function, $\varphi_\alpha$, usually described by HN type equations for all types of glassformers. Finally, the string model predicts that the temperature dependence of the average relaxation time $\tau_\beta$ for the JG β-relaxation is of Arrhenius type, and the relaxation function $\varphi_\beta$ is described by the Cole-Cole empirical equation [16,23].

An interesting prediction of the string model is that besides the crossover temperature $T_A$ ($\sim V'/k_B$) from high temperature Arrhenius to low temperature VFT behavior arising from inter-string interactions, another crossover temperature $T_L$ $\left(\sim V/2.5k_B\right)$ $> T_A$ may exist (Fig.1) caused by intra-string interactions, as shown in Eq.(1). This phenomenon was firstly discovered by Lunkenheimer et al [14], hence the subscript $L$ used here for this temperature. For $T > T_A$, the decrease of the relaxation time $\tau_a$ with increasing temperature is steeper than if the temperature dependence of $\tau_a$ were governed by Arrhenius behavior. This striking behavior, reflected as a red line in Fig.1, has been observed in 1-cyanoadamantane, an orientational glass or plastic crystal [14], but not in other glassformers presumably because they may be chemically unstable at the high temperatures at which this phenomenon occurs. Any Arrhenius fitting to $\tau_a$ in Fig.1, carried out in a narrow temperature range above $T_L$, gives a too small and non-physical Arrhenius relation pre-factor. The string model provides a reasonable interpretation of that non-physical quantity given in table I of Ref.14.

As a conclusion, comparison of theory with experiments shows that the string model provides a unified molecular level mechanism for the α- and JG β-relaxations in glass dynamics. We would like to point out the absence of the JG β-relaxation in some orientational glasses (so-called plastic crystals) [14], because the RM amplitude is relatively small while $\tau_\beta$ is close to $\tau_\alpha$. As a result, the two relaxation processes look like a single loss peak in the frequency domain. The relevant results will be described in separate publications [23].

ACKNOWLEDGMENTS

The authors thank Prof. E. Riande for his enlightening discussions. This work was supported by the National Natural Science Foundations of China (Grant No. 10774064 & 10274028), the Key Natural Science Foundation of Xinjiang Educational Department (2008-2010), the Key Natural Science Foundation of Xinjiang Science-Technology Department (2008-2010), and Key Natural Science Foundations of Yili Normal University (2006-2008 & 2007-2009).